\documentclass{ws-procs975x65b}

\begin{document}

\title{$B$ decays into $\pi\pi K$ and $K\bar KK$: long distance and final-state effects}


\author{B. Loiseau\footnote{\uppercase{I}nvited talk at the \uppercase{T}hird \uppercase{A}sia-\uppercase{P}acific \uppercase{C}onference on \uppercase{F}ew-\uppercase{B}ody \uppercase{P}roblems in \uppercase{P}hysics (\uppercase{APFB}05), \uppercase{N}akhon \uppercase{R}atchassima, \uppercase{T}ha\"{i}land, \uppercase{J}uly 26-30, to be publised by the \uppercase{S}cientific \uppercase{W}orld \uppercase{P}ublishing \uppercase{C}o. }}

\address{Laboratoire de Physique Nucl\'eaire et de Hautes 
\'Energies\footnote{\uppercase{U}nit\'e de \uppercase{R}echerche des \uppercase{U}niversit\'es
\uppercase{P}aris 6 et \uppercase{P}aris 7, associ\'ee au \uppercase{CNRS}}, Groupe Th\'eorie,\\
Univ. P. \& M. Curie, 4 Pl. Jussieu, F-75252 Paris, France\\ 
E-mail: loiseau@lpnhep.in2p3.fr}

\author{A. Furman, R. Kami\'nski, L. Le\'sniak}

\address{Department of Theoretical Physics,\\
The Henryk Niewodnicza\'nski Institute of Nuclear Physics,\\
Polish Academy of Sciences, 31-342 Krak\'ow, Poland\\
}  

\maketitle

\vspace{-0.4cm}

\abstracts{
The interplay of strong and weak decay amplitudes for $B\to \pi\pi K$ and $B\to K\bar KK$, with the $\pi\pi$ and $K\bar K$ pairs interacting in isospin-0 $S$-wave, is analyzed for $\pi\pi$ effective mass from threshold to 1.2 GeV.
To improve agreement with experiment of a factorization approach with some QCD corrections, addition of long-distance contributions, called charming penguins is necessary.}

\vspace{-0.6cm}

\section{Introduction}
Hadronic two- and three-body $B$-meson weak decays are a rich source of information to study CP violation within the Standard Model (and beyond). 
On Dalitz plots, many resonances are distinguishable but with entangled interference patterns.
Moreover the understanding of the final state interactions is weighty to attain accurate values of the Cabibbo-Kobayashi-Maskawa (CKM) matrix elements.
We report here on our recent study\cite{furm05} (FKLL) of $B\to \pi\pi K$ and $B\to K\bar KK$ decays.
We consider the $\pi\pi$ or $K\bar K$ pairs, denoted by $(\pi\pi)_S$ or $(K\bar K)_S$, to interact in isospin-0 $S$-wave in the $m_{\pi\pi}$ effective mass range from threshold to 1.2 GeV.
We extend to the $f_0(980)$ case the approach of Gardner and Mei\ss ner\cite{gard02} (GM) who have examined the effect of the $f_0(600)$ (or $\sigma$) on the $B^0\to \pi^+\pi^-\pi^0$ decays.
The decay amplitudes are calculated within the QCD factorization approach.
Due to cancellations between penguin amplitudes, the $B\to f_0(980)K$ branching fractions are too small compared to data.
We add long-distance contributions, called charming penguins.
Expressions for the  $B$ decay amplitudes and description of the final state interactions,  are given in section II. Comparison of our results to the data
and some conclusions are presented in section III.

\section{$B$ decay amplitudes and final state interactions}
From the effective weak Hamiltonian, $H_{eff}$, and the operator product expansion, the  $B$ decay amplitude into mesons $M_1$ and $M_2$ is\cite{baba98} ($l$ stands for quarks $d$ or $s$)
\begin{equation}
\label{BM1M2}
\langle M_1M_2\vert H_{eff}\vert B\rangle=
\frac{G_F}{\sqrt{2}}\ \sum_{q=u,c} V_{q,l}^{CKM}\left [\ \sum_{k=1}^{10} C_k(\mu)
\langle M_1M_2\vert O_k^{q,l}(\mu)\vert B\rangle ]\right ].
\end{equation}
In (1) $G_F=1.66\times 10^{-5}$ GeV$^{-2}$ is the Fermi constant, $V_{q,l}^{CKM}$ represents the CKM matrix-element factors and $\mu$ is a renormalization scale.
The perturbative Wilson coefficients $C_k(\mu)$ characterize the short-distance physics.
The non-perturbative hadronic matrix elements 
$\langle M_1M_2\vert O_k^{q,l}(\mu)\vert B\rangle$
describe the long-distance physics.
The local operators $ O_k^{q,l}(\mu)$ \textit{"effectively"} govern  the decay.
The main task of the theory is to compute these matrix elements in a reliable way. Beneke, Buchalla, Neubert and Sachrajda\cite{bene01} have developed the following QCD factorization formula,
\begin{equation}
\label{QCDFAC}
\langle M_1M_2\vert O_k^{q,l}(\mu)\vert B\rangle\sim
\langle M_1\vert J_1\vert 0\rangle 
\langle M_2\vert J_2\vert B\rangle
\left[
1+\sum_nr_n\alpha_s^n+O(\Lambda_{QCD}/m_b)
\right].
\end{equation}
In (2) $J_1$ and $J_2$ are bilinear quark currents, 
$\langle M_1\vert J_1\vert 0\rangle$ is the meson $M_1$ decay constant and 
$\langle M_2\vert J_2\vert B\rangle$ represents the transition form factor of the meson $B$ into the meson $M_2$.
The radiative corrections $r_n$ have been calculated to order $n=1$, $\alpha_s$ being the strong coupling constant.
The term $O(\Lambda_{QCD}/m_b)$ in Eq.~(\ref{QCDFAC}) represents power corrections and if set to zero,  together with $r_n$, one recovers the naive factorization formula as applied by Bauer, Stech and Wirbel\cite{baue87}.

The possible quark line diagrams for the 
$B^-\to(\pi\pi)_SK^-$ and
$B^-\to(K\bar K)_SK^-$
are shown in Fig.~1 of FKLL\cite{furm05}.
For the $B^0$ decays there is no tree diagram (a), only penguin diagrams similar to (b) or (c) diagrams exist.
Within the QCD factorization framework, just reminded above, the 
$B^-\to(\pi^+\pi^-)_SK^-$
decay amplitude is 
\begin{eqnarray}
\label{amplitude}
\langle (\pi^+\pi^-)_S K^-\vert H_{eff}\vert B^-\rangle & = & 
\frac{G_F}{\sqrt{2}}\sqrt{\frac{2}{3}}
\Big\{\chi
\left[
P(m_{\pi\pi})(U_a+U_b)+C(m_{\pi\pi})
\right]
\Gamma_1^{n*}(m_{\pi\pi})+
\nonumber \\
&&+ \left[
Q(m_{\pi\pi})U_c+ \chi C(m_k)
\right]\Gamma_1^{s*}(m_{\pi\pi})
\Big\}.
\end{eqnarray}
The constant $\chi$, to be fitted to reproduce the averaged branching fraction 
$\mathcal B[B^\pm\to f_0(980)K^\pm]$, is estimated to 30 GeV$^{-1}$ from the $f_0(980)$ decay properties\cite{furm05}.
The non-strange and strange pion scalar form factors $\Gamma_1^n(m_{\pi\pi})$ and 
$\Gamma_1^s(m_{\pi\pi})$ describe the final-state interactions. In (3)
$P(m_{\pi\pi})$ corresponds to the product of the kaon decay constant $f_K$ times the transition form factor of $B$ into $(\pi\pi)_S$, 
$F_0^{B\to (\pi\pi)_S}(M_K^2)$, 
\begin{equation}
\label{facfk}
P(m_{\pi\pi})=f_K(M_B^2-m_{\pi\pi}^2)F_0^{B\to(\pi\pi)_S}(M_K^2).
\end{equation}
The function $Q(m_{\pi\pi})$ is proportional to the product of the $(\pi\pi)_S$ "decay constant"  times the transition form factor of $B$ into $K$, $F_0^{B\to K}(m_{\pi\pi}^2)$, and reads\cite{furm05,gard02},
\begin{equation}
\label{facf0}
Q(m_{\pi\pi})=\frac{2\sqrt{2}B_0}{m_b-m_s}\ 
\left(
M_B^2-M_K^2
\right)
F_0^{B\to K}
\left(
m_{\pi\pi}^2
\right),
\end{equation}
with $B_0=-\langle 0\vert q\bar q\vert 0\rangle/f_\pi^2=m_\pi^2/(m_u+m_d)$. 

The functions $U_a,\ U_b$ and $U_c$, given in terms of the coefficients\cite{bene01,ali98,bene03,groo03} $a_i$ of the decay operators, are calculated from the Wilson coefficients.
They correspond to the contributions of Figs.~1(a), 1(b) and 1(c) of FKLL\cite{furm05}, respectively. 
One has
$U_a  =  V_{ub}V_{us}^*a_1,\ 
U_b  =  V_{ub}V_{us}^*\left(a_4^{u}-a_6^{u}r\right)+V_{cb}V_{cs}^* 
\left(a_4^{c}-a_6^{c}r\right)
 = V_{ub}V_{us}^*
\left[a_4^{u}-a_4^{c}+\left(a_6^{c}-a_6^{u}\right)r\right]
+V_{tb}V_{ts}^*\left(a_6^{c}r-a_4^{c}\right)\label{ub}$ and 
$U_c  =  -V_{ub}V_{us}^*a_6^{u}-V_{cb}V_{cs}^*a_6^{c}=
V_{ub}V_{us}^*\left(a_6^{c}-a_6^{u}\right)+V_{tb}V_{ts}^*a_6^{c}\label{uc}$.
We have used the unitarity of the CKM matrix elements, 
$V_{ub}V_{us}^*+V_{cb}V_{cs}^*+V_{tb}V_{ts}^*=0$.
In the reduction of the four-quark operators, split into product of two-matrix elements (Eq.~(\ref{QCDFAC})), a Fierz transformation is needed for the penguin diagrams in order to match
the flavor quantum numbers of the quark currents to  those of the hadrons\cite{ali98}.
The chiral factor $r$ is equal to
$2M_K^2/[(m_b+m_u)(m_s+m_u)]$.
Here and in Eq.~(\ref{facf0}), quark masses appear  as one makes use of the Dirac equation.
The $a_1$ and $a_{4,6}^{u,c}$ coefficients at next-to-leading order in $\alpha_s$ are given in Eq.~(35) of Beneke and Neubert\cite{bene03} (BN).
In the numerical calculations we take the values of table III of de Groot, Cottingham and Whittingham\cite{groo03} (GCW) at the scale $\mu=2.1$ GeV.
We do not include the small contributions from hard spectator interactions\cite{bene03}, annihilation and electroweak diagrams.

To improve their fit to charmless hadronic two-body $B$ decays with pseudoscalar-pseudoscalar and pseudoscalar-vector decays, GCW\cite{groo03}, following Ciuchini et al.\cite{ciuc97}, introduce the charming penguin term $C(m)$, the parametrization of which is reminded in Eq.~(6) of FKLL\cite{furm05}.
Expressions for the other $B\to(\pi\pi)_SK$ and for the $B\to(K\bar K)_SK$ amplitudes are given in FKLL\cite{furm05}.
If one neglects the penguin contraction terms (see Eq. (35), (39) and (41) of 
BN\cite{bene03}) then $a_{4,6}^{u}=a_{4,6}^{c}=a_{4,6}$ and our amplitude~(\ref{amplitude}) is similar to that of the $B^-\to\sigma\pi^-$ decay\cite{gard02} (see their Eq.~(25)).
Note the near cancellation\cite{furm05} in the $b\to s$ transition of the two penguin contributions as $r\sim 1$ and $a_4^c\sim a_6^c$.

\begin{figure}[!h]
\begin{center}
\includegraphics*[scale=0.40,angle=90]{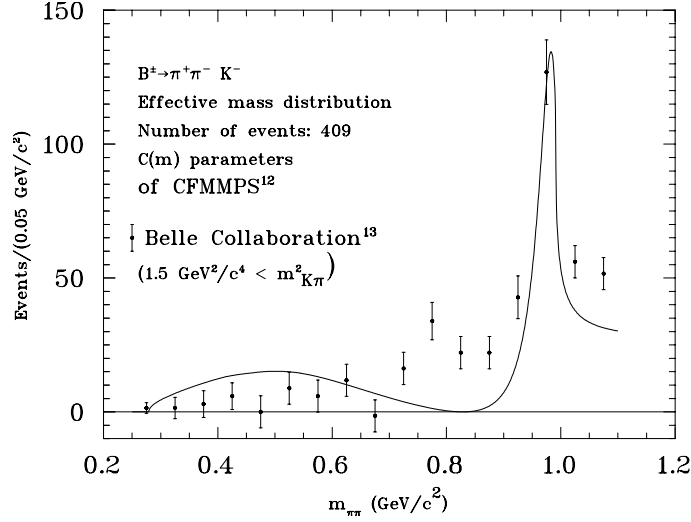}
\caption{Comparison of our model II with Belle data\protect\cite{bell0412}
\label{fig}}
\end{center}
\end{figure}

\begin{table}[!h]
\tbl{Average branching fractions $\mathcal{B}$ in units of $10^{-6}$, direct asymmetries 
$\mathcal{A}_{CP}$, $\mathcal{A}$ and $\mathcal{S}$ parameters of our model compared to data.
Model errors come from $C(m)$ parameters fitted by GCW\protect\cite{groo03}, model I, and CFMMPS\protect\cite{ciuc04}, model II. For a qualitative comparison we give in parenthesis the experimental  $B\to K\bar KK$ decay observables  obtained by considering the full range of the  $K\bar K$ invariant mass rather than the upper limit\protect\cite{furm05} of 1.1 GeV.
For data\protect\cite{baba0408076,bell0409} we only quote statistical errors.}
{\footnotesize
\begin{tabular}{ccccc}
\hline
	   &            & Average                      &  Model I       & Model II\\

\raisebox{1.5ex}[0pt][0pt]{$B$ decay mode} &  &  HFAG's values$^{15}$
	   &$\chi=33.5$ GeV$^{-1}$& $\chi=23.5$ GeV$^{-1}$\\
\hline
\hline\\[-7pt]
$B^\pm\to f_0(980)K^\pm$,    &$\mathcal{B}$   & $8.49^{+1.35}_{-1.26}$  & 8.49 (fit)  & 8.46 (fit) \\[2pt]
 $f_0\to \pi^+\pi^-$     &$\mathcal{A}_{CP}$& $-0.02\pm 0.07$   &$-0.52\pm 0.12$ & $0.20\pm 0.20$\\
\hline\\[-7pt]
 $B^0\to f_0(980)K^0$,   &$\mathcal{B}$     & $6.0\pm{1.6}$ &$5.9\pm 1.6$    & $5.8\pm 2.8$\\[2pt]
 $f_0\to \pi^+\pi^-$     &$\mathcal{A}$     & $-0.14\pm0.22$&$0.01\pm 0.10$  & $0.0004\pm 0.0010$\\[2pt]
         &$\mathcal{S}$     & $-0.39\pm0.26$&$-0.63\pm 0.09$ & $-0.77\pm 0.0004$\\
\hline\\[-7pt]
 \raisebox{-1.5ex}[0pt][0pt]{$B^\pm\to \left(K^+K^-\right)_SK^\pm $}         &$\mathcal{B}$   & $<2.9 ^{13}$& $1.8\pm 0.4$   & $1.7\pm 0.7$\\[2pt]
                                          &$\mathcal{A}_{CP}$& no data       &$-0.44\pm 0.12$ & $0.29\pm 0.21$\\
\hline\\[-7pt]
                                          &$\mathcal{B}$     & no data       &$1.1\pm 0.3$    &$1.2\pm 0.5$\\
 \raisebox{-1.5ex}[0pt][0pt]{$B^0\to \left(K^+K^-\right)_SK_S^0$}      &$\mathcal{A}$     & $(-0.09\pm0.10)$&$0.01\pm 0.10$  & $0.001\pm 0.001$\\[2pt]
            &\raisebox{-1.5ex}[0pt][0pt]{$\mathcal{S}$}     &
	    $(-0.55\pm0.22)^{14}$&\raisebox{-1.5ex}[0pt][0pt]{$-0.64\pm 0.09$} &
	    \raisebox{-1.5ex}[0pt][0pt]{$-0.77\pm 0.0006$}\\
                                          &     & $(-0.74\pm0.27)^{16}$& & \\
\hline\\[-7pt]                                         
                                          &$\mathcal{B}$     & no data       &$1.1\pm 0.3$    &$1.2\pm 0.5$\\
 $B^0\to \left(K^0_SK^0_S\right)_SK_S^0$  &$\mathcal{A}$     & $(0.41\pm0.21)$ &$0.01\pm 0.10$  & $0.001\pm 0.001$\\
                                          &$\mathcal{S}$     & $(-0.26\pm0.34)$&$-0.64\pm 0.09$ & $-0.77\pm 0.0006$\\
\hline
\end{tabular}\label{tab} }
\end{table}

The $\Gamma_1^{n,s}(m)$ obey unitary $\pi\pi$ and $K\bar K$ coupled channel equations\cite{furm05}.
We use the solution A of the two-body scattering matrix of Kami\'nski, Le\'sniak and Loiseau\cite{kami97}.
Below the $K\bar K$ threshold 
$\Gamma_1^{n,s^*}(m)=R_1^{n,s}(m)\cos\delta_{\pi\pi}(m)e^{i\delta_{\pi\pi}(m)}$.
$\vert\Gamma_1^{n,s}\vert$ is maximum  for $\delta_{\pi\pi}\simeq 180^{\circ}$, which corresponds to the $f_0(980)$ production.
The $R_1^{n,s}(m)$, depicting the formation of mesons prior to rescattering,
have been obtained by Mei\ss ner and Oller\cite {meis01}.

\section{Results and conclusions}
Our results with the charming-penguin $C(m)$ terms of GCW\cite{groo03}, model I, and of Ciuchini, Franco, Martinelli, Masiero, Pierini and 
Silvestrini\cite{ciuc04} (CFMMPS), model II, are compared to the 
data\cite{bell0412,baba0408076,hfag,bell0409} in Table 1.
Values of all parameters used are given in FKLL\cite{furm05}.
The effective $m_{\pi\pi}$ mass distribution for $B^\pm\to \pi^+\pi^-K^\pm$ of our model II is compared with Belle\cite{bell0412} in Fig. 1.
The sharp maximum near 1 GeV is from the $f_0(980)$.
Near 0.5 GeV the broad bump is related to the  $f_0(600)$.
Events near 0.8 GeV are due to the $B^\pm\to\rho^0K^\pm$ decay not considered in our model.
Similar agreement with $m_{\pi\pi}$ effective mass distributions of BaBar and Belle for charged and neutral B decays can be seen in the figures shown in FKLL\cite{furm05}.

To conclude, our effective mass distributions and branching fractions compare well with data.
The long distance contribution $C(m)$ is important.
If $C(m)=~0$, $\mathcal B(B^0\to\pi^+\pi^-K_S^0)$ is too small by a factor $\sim 18$ and 
$\mathcal B(B^\pm\to\pi^+\pi^-K^\pm)$ by a factor $\sim 4$.
If charming penguins are included the agreement with data is good with a $\chi$ compatible with the $f_0(980)$ properties.
The recent determination\cite{hfag} of $\mathcal{A}_{CP} = -0.02 \pm 0.07$  
for $B^\pm\to f_0(980)K^\pm,\ f_0\to \pi^+\pi^-$,  should help to determine better 
the charming-penguin $C(m)$ parameters.
In progress we are adding the $P$-wave contribution which will allow us to describe the 
$\rho$-meson region.

\section*{Acknowledgments}
We acknowledge good advices from B. El-Bennich.
This work benefits from IN2P3-Polish Laboratories Convention (project N$^{0}$ 99-97).

\end{document}